\documentclass[12pt]{article}

\usepackage[a4paper,left=2cm,right=2cm,top=1cm,
bottom=2cm]{geometry}

\usepackage{times}
\usepackage{amsmath}
\usepackage{siunitx}
\usepackage[T1]{fontenc}
\usepackage[english]{babel}
\usepackage[utf8]{inputenc}
\usepackage{epsfig}
\usepackage{epstopdf}
\usepackage{multirow}
\usepackage{graphicx}
\usepackage{caption}
\usepackage{url}
\usepackage{float}
\usepackage{array}
\usepackage{multirow}
\usepackage{amssymb}
\usepackage{caption}
\usepackage{subcaption}
\usepackage{titlesec}
\usepackage{cite}
\usepackage[table]{xcolor}
\usepackage{adjustbox} 
\usepackage{lipsum}
\usepackage[labelfont=bf]{caption}
\usepackage{color}
\usepackage{hyperref}
\hypersetup{
  colorlinks=true,
  linkcolor=blue,
  citecolor=blue,
  urlcolor=blue,
}
\usepackage{booktabs}

\usepackage[deletedmarkup=xout]{changes}

\definechangesauthor[color=cyan]{TM}

\definechangesauthor[color=blue]{IK}
\definechangesauthor[color=red]{TL}
\definechangesauthor[color=orange]{VM}
\definechangesauthor[color=orange]{DS}
\definecolor{cornflowerblue}{RGB}{100,149,237}
\definechangesauthor[color=red]{ZD}
\definechangesauthor[color=purple]{HP}

\makeatletter
\def\thickhline{%
  \noalign{\ifnum0=`}\fi\hrule \@height \thickarrayrulewidth \futurelet
   \reserved@a\@xthickhline}
\def\@xthickhline{\ifx\reserved@a\thickhline
               \vskip\doublerulesep
               \vskip-\thickarrayrulewidth
             \fi
      \ifnum0=`{\fi}}
\makeatother

\newenvironment{sciabstract}{%
\begin{quote} }
{\end{quote}}

\newcounter{lastnote}

\title{Synergistic control of radical generation in a radio frequency atmospheric pressure plasma jet via voltage waveform tailoring and structured electrodes }

\author
{M\'at\'e Vass$^{1,2\ast}$, Xiaokun Wang$^1$,
 Ihor Korolov$^1$,  Julian Schulze$^{1}$, Thomas Mussenbrock$^1$   \\
}

\date{}

\begin{document}

\maketitle 
 \vspace{-1cm}
 {\small
\begin{flushleft}

 $^{1}$ Chair of Applied Electrodynamics and Plasma Technology, Ruhr-University Bochum, 44780 Bochum, Germany \\
 
 $^{2}$  Institute for Solid State Physics and Optics, HUN-REN Wigner Research Centre for Physics, 1121 Budapest, Hungary\\

\vspace{0.5cm}
E-mail: vass@aept.rub.de
\end{flushleft}
}
\vspace{0.3cm}

\begin{sciabstract}
The synergy between voltage waveform tailoring and structured electrodes is investigated in a radio-frequency (RF) atmospheric-pressure microplasma jet operated in helium with a 0.1\% oxygen admixture. The device incorporates rectangular trenches in both electrodes and is driven by “Peaks” and “Valleys” waveforms synthesized from four harmonics (base frequency $f_{\rm b} = 13.56$~MHz, $V_{\rm pp} = 500$~V, $P=$1.2~W). Two-dimensional plasma fluid simulations, together with spatially and temporally resolved optical diagnostics (Phase-Resolved Optical Emission Spectroscopy and Tunable Diode Laser Absorption Spectroscopy), are used to demonstrate that the combination of asymmetric voltage waveforms with electrode structuring leads to strong spatial localization of electron power absorption and radical generation. This synergy results in a single pronounced maximum inside a trench at either the powered or grounded electrode, depending on the applied waveform, unlike a symmetric excitation, which produces a spatially symmetric enhancement at both electrodes. The effect is attributed to the interplay between waveform-induced sheath dynamics and geometric focusing provided by the trenches, enabling electrically reversible and selective enhancement of electron power absorption at a chosen location.
\end{sciabstract}

\section{Introduction}

Atmospheric-pressure, radio-frequency (RF) driven microplasma jets ($\mu$APPJs) have become valuable tools in plasma chemistry, largely owing to their pronounced nonequilibrium electron dynamics, which facilitate the formation of a diverse array of reactive species \cite{reuter2018kinpen, adamovich20222022}. This unique chemistry underpins a broad spectrum of applications, from plasma medicine and agriculture (where reactive oxygen and nitrogen species (RONS) are especially relevant) \cite{dharini2023cold, koga2024cold, nicol2020antibacterial}, to plasma-assisted synthesis \cite{stewig2023impact, schuttler2024production} and advanced surface processing \cite{dey2020engineering, rahman2024surface,shang2024atmospheric}.

The need for effective control and optimization of these devices is paramount, particularly regarding the selective manipulation of the electron energy distribution function (EEDF). One promising approach is Voltage Waveform Tailoring (VWT), in which a carefully designed excitation waveform composed of multiple harmonics can be used to control the high-energy tail of the EEDF and, consequently, influence plasma chemistry \cite{gibson2019disrupting, korolov2019control, korolov2021energy}. Significant progress has been made in both simulation \cite{liu2021micro, vass2021electron, vass2024energy} and experiment \cite{korolov2020helium, harris2024ozone, hubner2021effects}, demonstrating how the EEDF evolves in response to different excitation waveforms. In most cases, voltage waveforms with sharp peaks are used, leading to rapid sheath expansion and collapse near one electrode. This mechanism enhances electron power absorption in specific spatio-temporal regions, increases the population of high-energy electrons, and thus promotes the generation of radicals through electron-impact processes with high energy thresholds. Importantly, it has been demonstrated, both experimentally and in simulations, that VWT not only boosts the production of such radicals, but does so with greater energy efficiency. This has been observed in both He/N$_2$ \cite{korolov2019control} and He/O$_2$ \cite{vass2024energy} mixtures at atmospheric pressure. In the latter study, it was shown that while traditional “Peaks” and “Valleys” waveforms yield the highest mean densities of certain key radical species, the most effective waveform can depend on the specific radical targeted: sometimes, an alternative tailored voltage waveform offers greater selectivity or efficiency. Similar results were found in low pressure capacitively coupled RF discharges that are relevant for semiconductor manufacturing. Under such conditions high frequency VWT was found to provide enhanced EEDF control and more energy efficient production of radicals, e.g., F atoms, through tailoring the electron power absorption dynamics \cite{Derzsi_2013,Lafleur_2016,Wang_2021,Wang_2024}. The underlying mechanisms are, however, different from those at atmospheric pressure due to the different modes of electron power absorption.

Another approach for the control of radio-frequency discharges is the modification of the electrode topology, which has recently attracted significant interest at both low pressures \cite{zhang2025experimentalXiaokun, wang2021effects, wang2023effects} and intermediate pressures \cite{vdurian2022experimental, liang2024investigation, park2025two}, since such modifications can lead to a local increase in electron power absorption. Especially at low pressure, the main mechanism responsible for this effect is the radio-frequency Hollow Cathode Effect \cite{lee2010effective, lafleur2012particle}. In this scenario, if the plasma is able to penetrate into a trench, electrons accelerated by the expanding sheath at the trench walls can reach high energies, which enables them to participate in various inelastic processes, and in some cases become trapped by bouncing between the sheath edges. As the pressure increases and the electron energy relaxation length becomes shorter, this mechanism is somewhat modified. However, the underlying idea, that either bulk or secondary electrons can contribute to increased local power absorption, remains valid even at higher pressures \cite{park2025two}. This was also demonstrated by Liu et al. \cite{liu2023local}, who investigated the COST radio-frequency microplasma jet with trenched electrodes under single-frequency excitation in a He/O$_2$ mixture. They found that local maxima of electron power absorption occur near the trenches as a result of a current focusing effect: electrons accelerated by the locally expanding sheath edge are concentrated within a small spatio-temporal region, where they can efficiently drive inelastic processes. This local enhancement was found to increase the generation of certain radicals, such as helium metastables and atomic oxygen, indicating that electrode structuring can lead to a more efficient device design for the generation of reactive species.

In this paper, it is demonstrated through both simulations and experimental measurements that a clear synergy exists between voltage waveform tailoring and structured electrodes. This synergy enables precise control over radical generation by selectively manipulating the spatio-temporal regions of significant electron power absorption, which are responsible for electron-impact-driven formation of neutral species. The paper is organized as follows: section~\ref{sec:exp} describes the simulation and experimental methods; section~\ref{sec:res} presents the results; and section~\ref{sec:conc} provides the conclusions.

\section{Experimental and computational method}\label{sec:exp}

In this work, we investigate a modified version of the COST microplasma jet, as described in detail in \cite{liu2023local}. The two electrodes, originally separated by 1 mm to define an active plasma volume of $1~{\rm mm} \times 1~{\rm mm} \times 30~{\rm mm}$, are further modified with rectangular trenches (width 0.5~mm, depth 1~mm, spaced 5~mm apart), which locally increase the volume available for plasma generation. The computational study employs the two-dimensional plasma fluid model {\it nonPDPSIM} \cite{norberg2015formation}, which solves the continuity equations for charged and neutral species (using the drift-diffusion approximation), Poisson’s equation, and the material charge density equation. The equations for charged species and electric potential are fully coupled and solved using a fully implicit Newton–Raphson iterative scheme. The neutral species equations are solved separately at each time step, with a time-slicing strategy used \cite{kushner2009hybrid} to advance them over longer intervals while plasma parameters are held fixed. For electrons, the local mean energy approximation is applied, and transport coefficients are determined from a two-term Boltzmann solver. The gas flow is modeled with a compressible Navier–Stokes solver, yielding the flow field, temperature, and fluid density. The model utilizes a triangular unstructured mesh.

\begin{figure}[H]
    \centering
    \includegraphics[width=0.95\linewidth]{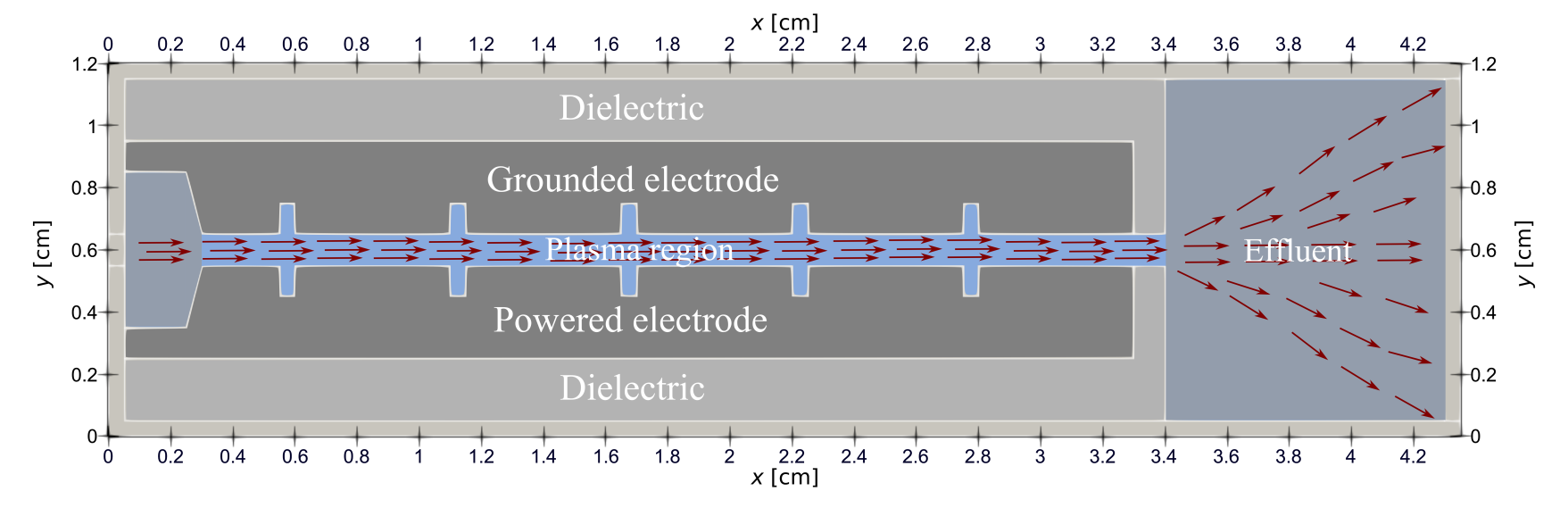}
    \caption{Schematic of the simulated two-dimensional domain. The grounded and powered electrodes, each modified with rectangular trenches (0.5 mm wide and 1 mm deep, spaced 5 mm apart), are enclosed by dielectric regions. The area shaded in blue indicates the plasma region. Red arrows illustrate the direction of the gas flow, shown for illustration purposes only. }
    \label{fig:mesh}
\end{figure}

The computational domain is illustrated in Fig.~\ref{fig:mesh}. The dielectric surrounding the electrodes is assigned a relative permittivity of $\epsilon_{\rm r}=4$, reflecting the use of quartz glass in the experimental setup. The trenches are 0.5 mm wide, which is sufficiently large to allow plasma penetration, yet not so wide that the current focusing effect is lost \cite{liu2023local}.

The flow field, depicted by red arrows in the figure, is shown for illustration only and is not based on the simulated velocity field; within the plasma region, the actual flow is laminar and closely approaches the Hagen–Poiseuille limit. The unstructured mesh comprises approximately 34,000 nodes, with 21,000 nodes located in the plasma region. The smallest resolved distance in the simulation is $3\times10^{-5}$~m. The model uses adaptive time-stepping for charged particles, with the timestep controlled by the number of Newton–Raphson iterations and an average value of $10^{-10}$~s. To ensure full convergence, the plasma was simulated for $60~\mu$s, while the neutrals and flow field were advanced for $0.1$~s.

\begin{figure}[H]
    \centering
\includegraphics[width=0.55\linewidth]{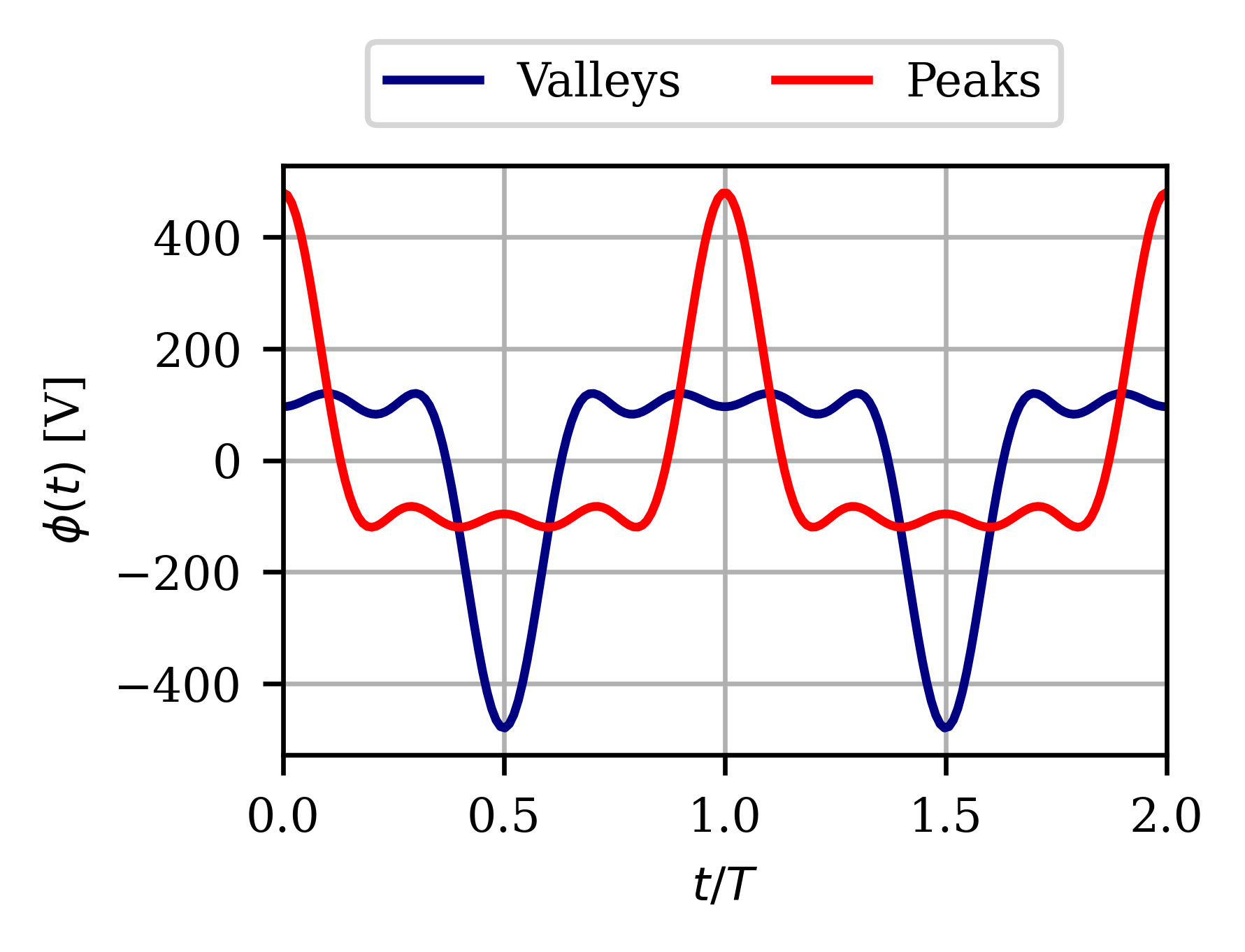}
    \caption{``Valleys'' and ``Peaks'' waveforms as a function of time over two fundamental RF periods.}
    \label{fig:P}
\end{figure}

The charged species considered in the He/O$_2$ model are electrons (e$^-$), positive ions (O$_2^+$, O$^+$, He$^+$), and negative ions (O$^-$, O$_2^-$, O$_3^-$). The neutral species include He, O$_2$, O, O$_3$, O$_2$($v$=1–4), O$_3$($v$), O$_2$(a$^{1}\Delta{_\mathrm{g}}$), O$_2$(b$^{1}\Sigma{_\mathrm{g}}^{+}$), O($^1$D), He(2$^3$S), and He(2$^1$S). The chemistry set and cross sections for the Boltzmann solver, as well as surface coefficients (including electron reflection, secondary electron emission from positive ions, and sticking coefficients for neutrals), are identical to those reported in \cite{liu2023local}.

For the same system, spatially resolved phase-resolved optical emission spectroscopy (PROES) was performed to measure excitation rates from the ground state to the He (3s) $^3$S$_1$ state, together with tunable diode laser absorption spectroscopy (TDLAS) for the He$(2^3$S) state density. Further details of these diagnostic methods can be found in \cite{liu2021micro}.

The gas mixture consisted of 99.9\% He and 0.1\% O$_2$. For both experiments and simulations, the ``Peaks'' and ``Valleys'' waveforms shown in Fig.~\ref{fig:P} were used; these were synthesized from four consecutive harmonics with a base frequency of $f_{\rm b}=13.56$~MHz. In the experiments, a peak-to-peak voltage of $V_{\rm pp}=500$~V was applied, corresponding to a total plasma power of 1.2 W. To achieve comparable power in the simulation, a higher peak-to-peak voltage of $V_{\rm pp}=600$~V was required. This difference arises from the relatively low oxygen content, which leads to an appreciable energy relaxation length and can result in deviations when using pure fluid models, even at atmospheric pressure \cite{vass2024new}.

\section{Results}
\label{sec:res}

In this section, both simulation and experimental results are presented for the COST jet with trenched electrodes excited by tailored voltage waveforms.

\begin{figure}[H]
    \centering
    \includegraphics[width=.87\linewidth]{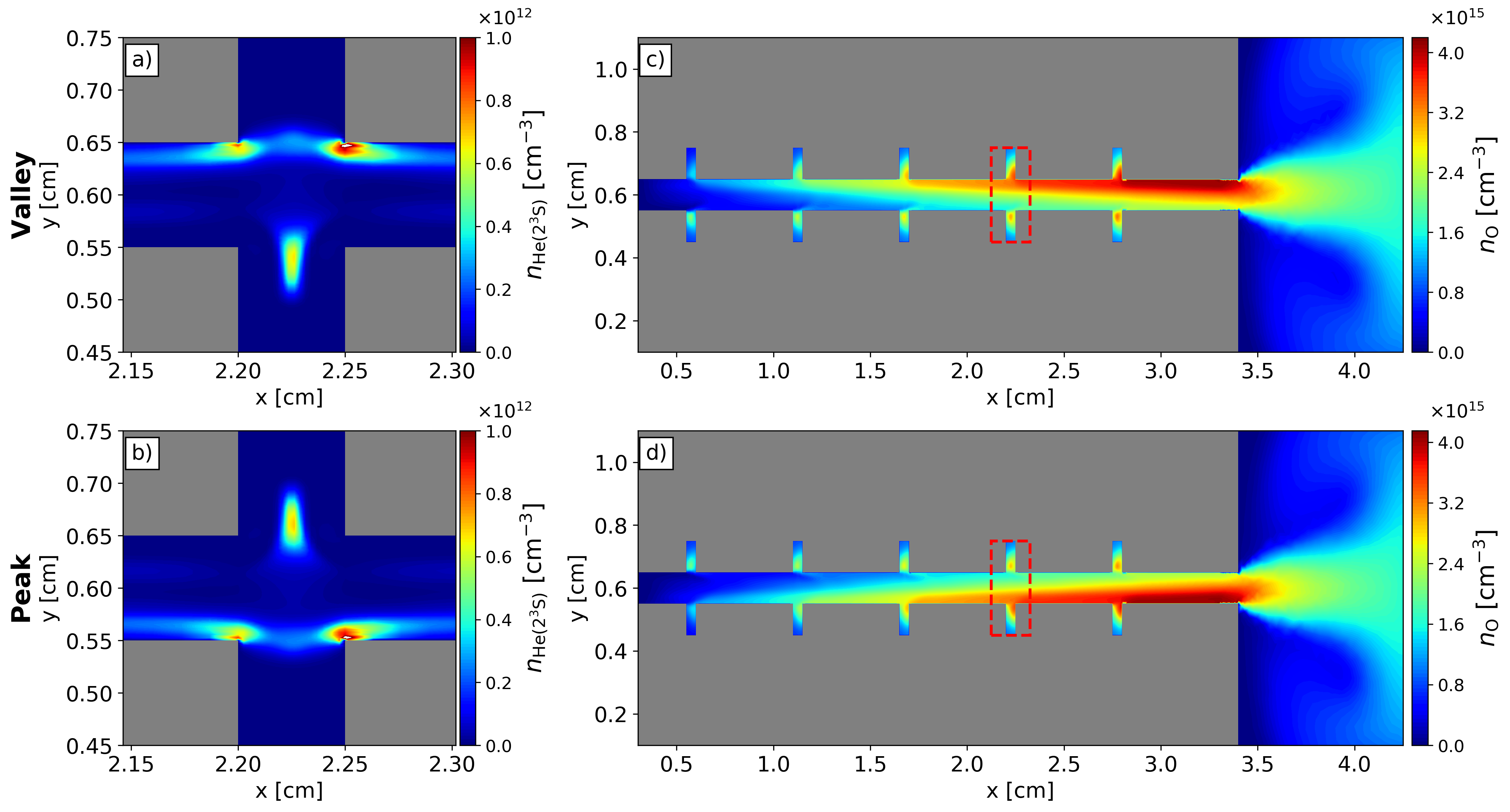}
    \caption{Simulation results for the time-averaged helium metastable density, $n_{\rm He(2^3S)}$ (left) and atomic oxygen density, $n_{\rm O}$ (right) shown for “Valleys” and “Peaks” types of driving waveforms (top/bottom rows, respectively). The red dashed rectangles in panels (c, d) indicate the trench region visualized in panels (a, b). Operating conditions: $p = 10^5$~Pa, $V_{\rm pp} = 600$~V, $f_{\rm b} = 13.56$~MHz, He/O$_2$~=~99.9/0.1.}
    \label{fig:valpeak}
\end{figure}

Figure~\ref{fig:valpeak} shows simulation results for the “Valleys” (top row) and “Peaks” (bottom row) waveforms, specifically highlighting the two-dimensional spatial distributions of the time-averaged helium metastable density, $n_{\rm He(2^3S)}$ (panels a,~b), and atomic oxygen density, $n_{\rm O}$ (panels c,~d). The red dashed rectangles in panels (c) and (d) indicate the trench region visualized in panels (a) and (b). As discussed in \cite{liu2023local,vass2024new}, the dynamics of helium metastables are primarily governed by electron impact excitation, which occurs on a much shorter timescale than chemical reactions involving neutral species. Consequently, $n_{\rm He(2^3S)}$ shows little spatial variation along the flow direction, allowing analysis to focus on a single trench. In contrast, the atomic oxygen density, which is strongly influenced by chemical reactions with other neutrals such as He, O$_2$, and O$_3$, increases markedly along the flow, as shown in panels (c) and (d).

Comparing panels (a) and (b), it is evident that the spatial excitation pattern can be controlled by the choice of “Valleys” or “Peaks” waveforms due to the induced electrical asymmetry. For the “Valleys” waveform, two local maxima in the helium metastable density appear near the grounded electrode at the trench edges, with an additional maximum inside the trench of the powered electrode. This arises from the interplay between the excitation waveform and the electrode topology. In the absence of a trench, the “Valleys” waveform causes the sheath near the grounded electrode to remain close to its full expansion for most of the RF cycle. In this region, electrons (either produced by ion-induced secondary emission or via Penning ionization from helium metastables) can acquire high energies, driving ionization and other high-threshold inelastic processes, particularly the formation of helium metastables (with a threshold of approximately 20 eV) \cite{vass2021electron}. This explains both the elevated helium metastable density in the region without trenches (e.g., near $x=2.16$~cm) and the highest atomic oxygen concentration near the grounded electrode, as shown in panel (c). The presence of a trench introduces two additional effects: (i) increased species generation near the trench edges due to the field enhancement at these geometrical features, and (ii) a current-focusing effect. As seen, for example, in panel (a), there is an asymmetry between the trench edges, with a higher helium metastable density observed at the edge aligned with the gas flow. This is attributed to the increased generation of atomic oxygen, which can contribute to electron production from negative ions (see, e.g., \cite{vass2024new}), ultimately resulting in enhanced excitation. As reported in \cite{liu2023local} for single-frequency excitation, expansion of the local sheath near a trench accelerates electrons, leading to a marked increase in electron power absorption within that trench. While single-frequency excitation produces symmetric sheath dynamics at both electrodes, the asymmetric “Valleys” and “Peaks” waveforms preferentially enhance {excitation at} one electrode, as shown in Fig. \ref{fig:valpeak} (a), and result in a locally elevated atomic oxygen density (see panel (c)). This demonstrates the synergy between electrode structuring and tailored voltage waveforms.

\begin{figure}[H]
    \centering
\includegraphics[width=0.95\linewidth]{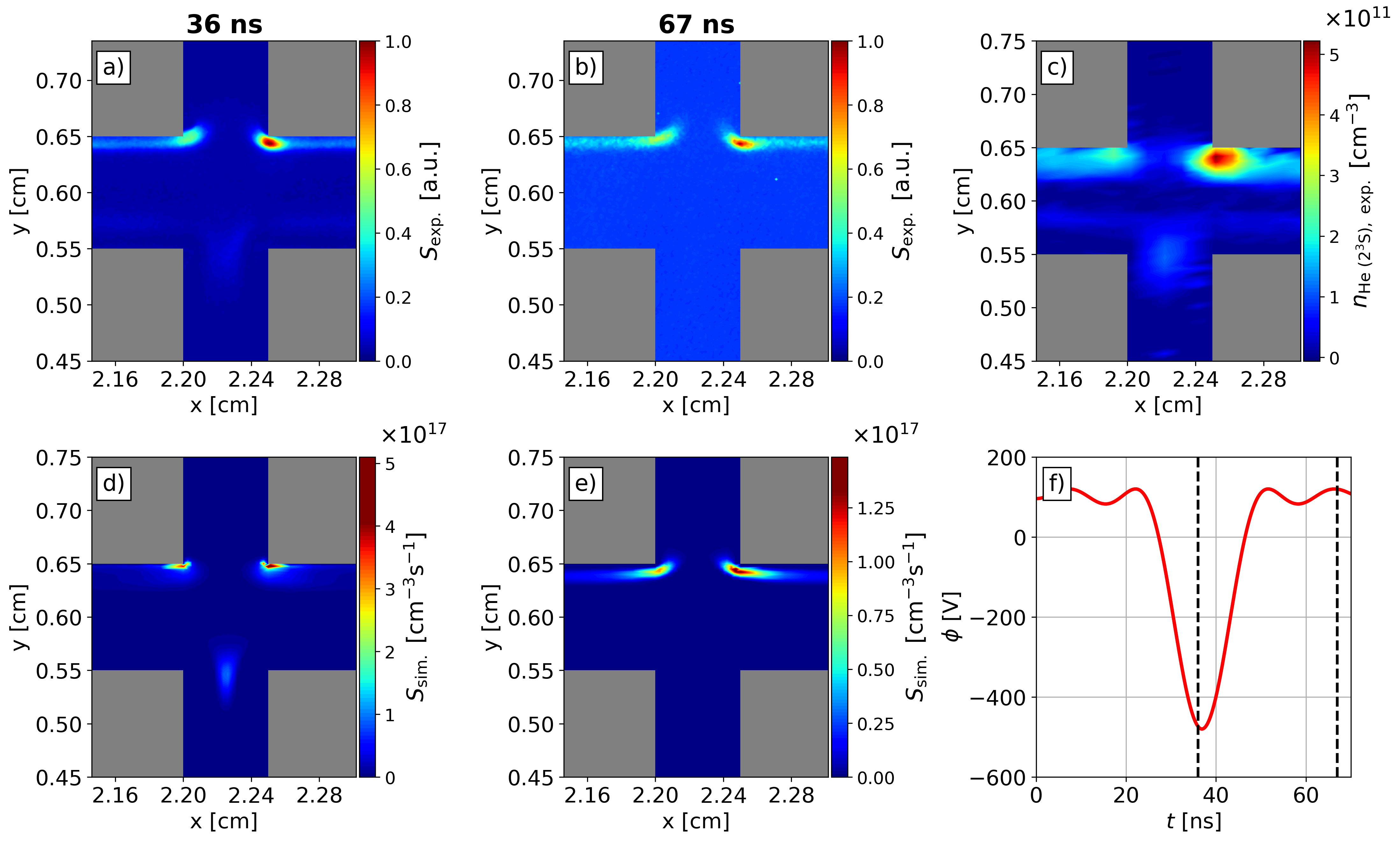}
    \caption{Experimental and simulated results of the excitation rate from the ground state to the He I (3s) $^3$S$_1$ state at 36~ns (a, d) and 67~ns (b, e), time-averaged TDLAS measurement of the helium metastable density $n_{\rm He\,(2^3S)}$ (c, cf.~Fig.~\ref{fig:valpeak}~(a)), and the applied excitation waveform (f), with the two time instances marked by vertical dashed lines. Operating conditions: $p = 10^5$~Pa, $V_{\rm pp} = 600$~V, $f_{\rm b} = 13.56$~MHz, He/O$_2$ = 99.9/0.1.}
    \label{fig:exp}
\end{figure}

To elucidate the underlying physics, we examine the “Valleys” waveform case in greater detail. Good qualitative agreement is observed between simulation results and experiments, as shown in Fig.~\ref{fig:exp}. This figure presents PROES measurements of the electron excitation rate from the ground state to the He I (3s) $^3$S$_1$ state at two specific time instances within the RF cycle (panels a and b), alongside the corresponding simulation results (panels (d) and (e)). Panel (c) shows TDLAS measurements of the helium metastable density. The applied excitation waveform is illustrated in panel (f), with two vertical dashed lines indicating the specific time points corresponding to panels (a), (b), (d), and (e). 

At $t=36$~ns, close to the voltage peak of the ``Valleys'' waveform, where the local sheath edge at the powered electrode is close to its full expansion, the current focussing effect is evident in both the experiment (panel (a)) and simulation (panel (d)) near the powered electrode, seen as a local increase in the excitation rate within the trench, although the relative magnitudes of the local maxima near the powered and grounded electrodes slightly differ between experiment and simulation. As seen in panels (b) and (e), at times when the sheath at the grounded electrode is near full expansion, the current focusing effect near the grounded electrode is absent, and high excitation is observed near the grounded electrode, particularly at the trench edges, as discussed above. The time-averaged experimentally measured helium metastable density (panel (c)) reflects both features, in alignment with Fig. \ref{fig:valpeak}(a): the current focusing effect produces a local maximum inside the trench at the powered electrode, while additional maxima occur near the grounded electrode and trench edges due to the electrically asymmetric waveform. In terms of absolute values, a factor of two discrepancy exists between measured and simulated excitation rates. This is expected when using voltage waveform tailoring in the COST jet at low molecular gas concentrations, as the resulting electron energy distribution functions are challenging to reproduce accurately with a fluid model employing a two-term Boltzmann solver \cite{vass2024energy}. 

\begin{figure}[H]
    \centering
    \includegraphics[width=\linewidth]{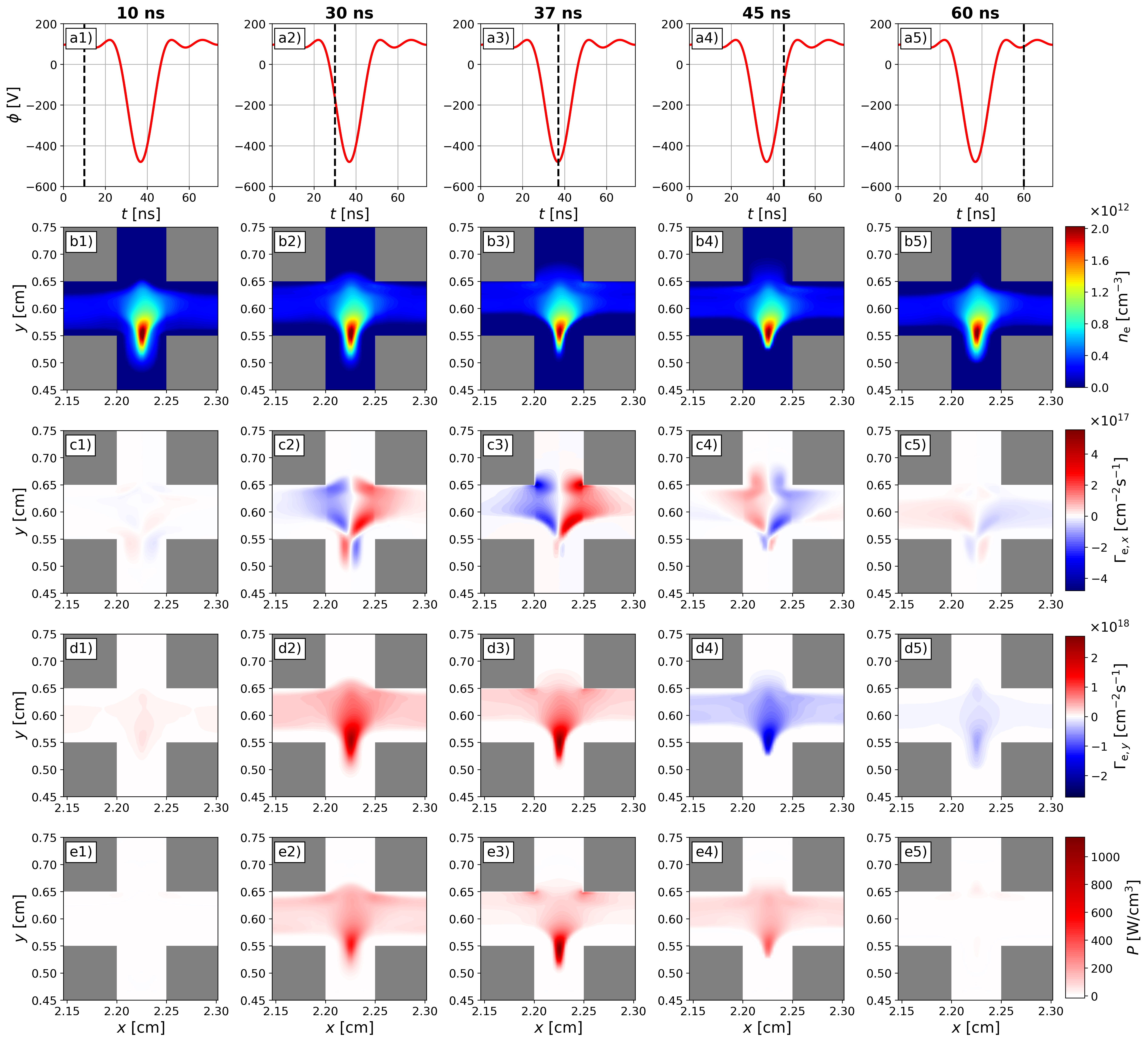}
    \caption{Driving voltage waveform (``valleys'' shape), $\phi(t)$ (first row); electron density, $n_{\rm e}$ (second row); $x$- and $y$-components of the electron flux, $\boldsymbol{\Gamma}_{\rm e}$ (third and fourth rows); and electron power absorption, $P$ (fifth row) are shown at selected time instances within one RF cycle. These times are indicated by vertical dashed lines in panels (a1–a5) and provided in the titles of the corresponding columns. Operating conditions: $p = 10^5$~Pa, $V_{\rm pp} = 600$~V, $f_{\rm b} = 13.56$~MHz, He/O$_2$~=~99.9/0.1.
  }
    \label{fig:array}
\end{figure}

To understand the physical basis for the synergy between voltage waveform tailoring and electrode topology, Fig. \ref{fig:array} presents simulation results for the spatial distribution of various physical quantities in the region near the trench, also shown in Fig. \ref{fig:valpeak}, at different times within the fundamental RF cycle. Panels (a1)–(a5) show the driving voltage waveform, with the time instances marked by vertical dashed black lines. In the figure, each column corresponds to one time instance, and each row displays a different physical quantity:  electron density $n_{\rm e}$ (panels (b1)–(b5)), the $x$- and $y$-components of the electron flux $\boldsymbol{\Gamma}_{\rm e}$ (panels (c1)–(c5) and (d1)–(d5), respectively), and the electron power absorption $P = -e\boldsymbol{\Gamma}_{\rm e} \cdot \mathbf{E}$ (panels (e1)–(e5)), where $\mathbf{E}$ is the electric field. Each column corresponds to a specific time selected from the waveform in panels (a1)–(a5). Due to the pronounced separation in timescales between electron dynamics and chemistry/gas flow, it is sufficient to focus on a single trench, rather than the entire jet, to elucidate the key characteristics of the observed synergy. At times when the sheath is close to its full expansion near the grounded electrode, such as in the first and fifth columns, both the electron flux and the electron power absorption are very small. This behavior can be attributed to the characteristics of the “Valleys” waveform: appreciable electron current only occurs during the period around the voltage peak, since it is primarily needed to compensate for the continuous ion flux arriving at the electrodes. As the sheath near the powered electrode starts to expand (second column), a pronounced current focusing effect emerges. This can be observed clearly in panels (c2) and (d2). The expanding sheath causes electrons to move upward, i.e., toward the grounded electrode. At the same time, the presence of the rectangular trench leads to a spatial focusing of the electron flux into a sharply defined region near the center of the trench, as seen in panel (c2). This geometrical focusing effect enhances the local electron density and results in a corresponding increase in power absorption, with a local maximum at the point of strongest current focusing. This enhanced power deposition in a very specific location is a direct manifestation of the interplay between the applied voltage waveform and the trench topology.

During the phase when the sheath is fully expanded at the powered electrode, corresponding to the minimum of the voltage peak (third column), some electrons can enter the trench at the grounded electrode. This results in elevated power absorption near the trench edges, as seen in panel (e3), particularly around $y = 0.65$~cm. However, at this stage, the electron flux inside the trench exhibits what can be called an “antifocusing” effect (see panels (c3) and (d3)): rather than being concentrated, the electron current becomes more dispersed within the trench as the electrons move toward the electrode due to sheath collapse. This observation is consistent with the results reported in \cite{liu2023local}, where it was shown that, under symmetric single-frequency excitation, the focusing and antifocusing phases occur with equal magnitude but opposite polarity, resulting in a symmetric distribution of electron density and power absorption inside the trench. In contrast, the asymmetry inherent in the “Valleys” voltage waveform used in this study breaks this symmetry. Consequently, only a single pronounced electron density peak and a corresponding maximum in power absorption are observed within the trench, both situated near the powered electrode, as shown in panels (b1) to (b5).

 This can also be seen in panel (c4), where a current focusing effect appears inside the trench at the grounded electrode, although it is less pronounced than at the powered electrode. This clear asymmetry in the voltage waveform is a fundamental factor behind the observed synergy between voltage waveform tailoring and electrode topology, as it enables spatial localization and control of plasma properties within specific regions of the device. These results demonstrate that combining voltage waveform tailoring with structured electrodes significantly enhances radical generation due to localized current focusing and increased electron power absorption near the trenches. This synergy allows for more energy-efficient plasma operation and offers a clear pathway toward optimizing device structures for advanced plasma processing applications.

\section{Conclusions}\label{sec:conc}

In this study, we have investigated the interplay between voltage waveform tailoring and electrode topology in an atmospheric-pressure RF microplasma jet operated in a He/O$_2$ (99.9/0.1) mixture. Using a combination of two-dimensional plasma fluid simulations and spatially and temporally resolved experimental diagnostics (PROES as well as TDLAS), we demonstrated how the electron power absorption, and thus the spatial distribution of radical generation, can be precisely manipulated by engineering both the electrical excitation and the electrode geometry. Our results show a clear synergy between waveform asymmetry and geometric structuring. When a “Valleys” type voltage waveform is applied to untrenched electrodes, radical generation is preferentially enhanced near the grounded electrode due to the asymmetric sheath dynamics. For single-frequency excitation in trenched electrodes, the current focusing effect leads to elevated excitation within the trenches at both electrodes, yielding a more symmetric pattern. However, when rectangular trenches are combined with an asymmetric tailored waveform, such as “Valleys,” the resulting sheath dynamics lead to a distinct spatial asymmetry: a single pronounced maximum in electron power absorption and radical generation appears inside the trench at the powered electrode, while two additional local maxima form near the grounded electrode at the trench edges. This behavior results from the interplay between the trench geometry and the asymmetric sheath dynamics, where the sheath remains fully expanded near the grounded electrode for most of the RF cycle, enabling continuous electron acceleration and enhanced excitation in that region. Furthermore, we have shown that the location of radical generation can be actively controlled by switching between “Valleys” and “Peaks” waveforms, effectively moving the region of enhanced power absorption between opposing trenches. This dynamic spatial control, confirmed by both simulation and experiment, is a direct manifestation of the synergy between electrical and geometric design. The insights gained here have broader implications for the optimization and control of plasma processes. By leveraging the combination of tailored voltage waveforms and structured electrodes, it is possible to enhance the selectivity and spatial localization of plasma-generated radicals.

\section*{Acknowledgements}

 This work was supported by the German Research Foundation in the frame of the collaborative research center SFB 1316, Project A4 and A5 and the DFG research project MU 2332/12-1. The authors thank Prof. Mark Kushner for providing the {\it nonPDPSIM} code.

\bibliographystyle{iopart-num-long}
\bibliography{main}

\end{document}